# Coexistence of light and heavy carriers associated with superconductivity and anti-ferromagnetism in CeNi$_{0.8}$Bi$_2$ with a Bi square net

Hiroshi Mizoguchi,[1] Satoru Matsuishi,[2] Masahiro Hirano,[1] Makoto Tachibana,[3] Eiji Takayama-Muromachi,[3] Hitoshi Kawaji,[2] and Hideo Hosono[1,2]

[1]Frontier Research Center, Tokyo Institute of Technology, 4259 Nagatsuta, Midori-ku, Yokohama 226-8503, Japan,

[2]Materials and Structures Laboratory, Tokyo Institute of Technology, 4259 Nagatsuta, Midori-ku, Yokohama 226-8503, Japan,

[3]National Institute for Materials Science, 1-1 Namiki, Tsukuba, Ibaraki 305-0044, Japan

Corresponding author footnote: hosono@lucid.msl.titech.ac.jp

**Abstract**

We found that the ZrCuSiAs-type crystal $CeNi_{0.8}Bi_2$ with a layered structure composed of alternate stacking of $[CeNi_xBi(1)]^{\delta+}$ and $Bi(2)^{\delta-}$ exhibits a superconductive transition at ~4 K. The conductivities, magnetic susceptibilities, and heat capacities measurements indicate the presence of two types of carriers with notable different masses, *i.e.*, a light electron responsible for superconductivity and a heavy electron interacting with the Ce 4f electron. This observation suggests that 6p electrons of Bi(2) forming the square net and electrons in $CeNi_xBi(1)$ layers primarily correspond to the light and heavy electrons, respectively.

Significant attention has been paid towards ZrCuSiAs-type compounds [1-10] since the discovery of superconductivity in F-doped LaFeAsO with a transition temperature ($T_c$) of 26 K. Maintaining the basic structural unit of FeAs, several new systems, such as LiFeAs (111), $AFe_2As_2$ (A = Sr and Ba) (122) and $FeSe_{1-x}$ (11) have been found to exhibit superconductivity, which coexists with an anti-ferromagnetic (AFM) state at low temperature. A large variety of superconductive compounds belonging to these layered crystal structures [11] provide a good platform to explore novel superconductors and to clarify the high $T_c$ mechanism. ZrCuSiAs-type compounds contain two types of anions. There are materials in which the same types of anions occupy the anionic sites. For example, $RNi_xBi_2$ (R = Ce, Nd, Gd, Tb, Dy and Y) contains two types of Bi ions: Bi(1) forms $NiBi_4$ and Bi(2) forms a Bi square net and the formal charge of the former is −3, whereas the latter has a charge of −1 [12,13]. **Figure 1** shows the crystal structure of $CeNi_xBi_2$, revealed by powder neutron structure analysis [14] in comparison with LaFeAsO [1]. Ni ions (the formal valence state is +1) with a stoichiometric composition deficiency occupy tetrahedral sites, and form a distorted tetrahedron with Bi(1) ions with a formal charge state of −3. Taking into consideration the ionic sizes of Ni and Ce, Bi(2) with the formal charge state of −1 occupies a narrower space, resulting in a shorter Bi(2)−Bi(2) spacing of 3.21 Å, which arises from the covalent nature of the bond (Bi 6p-Bi 6p σ/π bonding). This Bi-array is called the "Bi(2) square net"[11,15], where the Bi 6p-band is not fully occupied because $Bi^{-1}$ has an electronic configuration of $(6s)^2(6p)^4$. The higher valence state of the Bi(2) makes the $Ce-Bi(2)^{-1}$ distance (3.45 Å) longer than $Ce-Bi(1)^{-3}$ (3.34 Å), and the Ce ion forms tighter bonds with the more negative Bi(1) ion. The two-dimensional electronic structure of the Bi(2) square net overlaps with the $Ni_xBi(1)$ conductive layer near the Fermi level ($E_F$). In this letter, we report that $CeNi_xBi_2$ exhibits transitions due to both superconductivity at ~4 K and AFM at ~5 K, and the light electron of Bi(2) is responsible for the superconductivity, whereas the heavy electron is responsible for the strong interaction with Ce 4f which yields the AFM transition.

According to references 12 and 13, there exists distinct off-stoichiometry of Ni in this series of compounds. Thus, the optimal Ni content was determined in an effort to minimize the impurity phase identified by power X-ray diffraction. Polycrystalline samples of RNi$_x$Bi$_2$ (R = La, Ce, Nd and Y) were synthesized by solid-state reactions at elevated temperatures in evacuated silica ampoules. Starting materials used were La (99.9%), Ce (99.9%), Nd (99.9%), Y (99.9%), Ni (99.9%) and Bi (99.9%). Appropriate amounts of these mixtures corresponding to the optimal x value were heated in evacuated silica ampoule at 773 K for 10 h, followed by heat-treatment at 1023 K for 20 h. The products obtained were ground and pressed into pellets and an additional heat-treatment was performed in evacuated silica ampoules at 1073 K for 10 h. All the treatments of the starting materials were performed in an Ar-filled glove box (O$_2$, H$_2$O < 1 ppm). The crystal structure of the synthesized materials was examined by powder X-ray diffraction (XRD; Bruker D8 Advance TXS) using Cu K$_\alpha$ radiation with the aid of RIETVELD refinement using Code TOPAS3 [16]. The temperature dependence of the DC electrical resistivity ($\rho$) at 2–300 K was measured using a conventional four-probe method using Ag paste as the electrodes. Magnetization (M) measurements were performed with a vibrating sample magnetometer (Quantum Design). Specific heat data were obtained using a conventional thermal relaxation method using Quantum Design PPMS.

Sintered RNi$_x$Bi$_2$ polycrystalline samples with a dark gray color and metallic luster, decompose gradually into powder when exposed to an ambient atmosphere. Thus, the samples were stored in evacuated desiccators before measurements. The structure was confirmed to be ZrCuSiAs-type by powder XRD measurements. For the Ce compound, inductively coupled plasma spectroscopy was employed to confirm the chemical composition, CeNi$_{0.8}$Bi$_2$, which was consistent with the result obtained by the Rietveld refinement to the XRD patterns as well as the batch composition. **Table I** summarizes the Ni content (x), lattice constants and unit cell volume for RNi$_x$Bi$_2$. Although the Ni content fluctuates with the R ion, the unit cell volumes changed monotonically with the atomic number of the R ion, according to the lanthanide contraction rule, suggesting that the Ce ion takes a +3 charge state in CeNi$_{0.8}$Bi$_2$. **Fig. 2(a)** shows the resistivity-temperature ($\rho$-T) curve for CeNi$_{0.8}$Bi$_2$ under an applied magnetic field of 0 Oe. The resistivity is almost independent of T above 100 K, but decreases almost linearly with T at T < 100 K. With a further decrease in T, CeNi$_{0.8}$Bi$_2$ shows a T$^2$-dependence below 10 K. Although the absolute $\rho$-value varies with samples by an order of magnitude, the overall features remain unchanged. As shown in the inset of **Fig. 2(a)**, a sharp drop in $\rho$ was observed at T = 4.2 K, and the resistivity vanished at 4.0 K. The T$_c$ shifts to a lower side with increasing H, suggesting that the CeNi$_{0.8}$Bi$_2$ undergoes a superconducting transition at 4.2 K. The magnetic susceptibilities ($\chi$) measured in zero-field cooling processes (ZFC) reached −5.0 emu/mol-Ce (**Fig. 2(b)**). This value corresponds to the volume fraction of the superconductivity phase of 96% (estimated from the $\chi$ value of perfect diamagnetism), which confirms that the bulk superconductive transition takes place at 4.2 K. A small hump is also observed at ~7 K, as shown in the right inset. The M-H curve at 2 K in the left inset of **Fig. 2(b)** shows a typical profile for a type-II superconductor with a lower superconducting critical magnetic field (H$_{c1}$) of ~65 Oe. **Table II** summarizes T$_c$, H$_c$ and the superconducting volume fraction of four kinds of RNi$_x$Bi$_2$ compounds. Although T$_c$ is similar for the four compounds, the volume fraction of the superconductivity phase was found to be significantly enhanced in the Ce compound.

**Fig. 3** shows the heat capacity of CeNi$_{0.8}$Bi$_2$ as a function of temperature. A distinct $\lambda$ peak is observed only for CeNi$_{0.8}$Bi$_2$. Application of the magnetic field suppressed the $\lambda$ peak. The Sommerfeld coefficient ($\gamma$) at 5.0 K was estimated to be 0.4 J K$^{-2}$mol-Ce$^{-1}$ (inset of **Fig. 3**), which in turn led to an observable large mass-enhancement of the conducting carrier, because the $\gamma$ value is proportional to the density of states at E$_F$. Such a large mass enhancement was not seen for the La- or Y-systems. We thus attributed the observed $\lambda$ peak to AFM ordering of the Ce 4f electron spin, rather than the superconductive transition on the basis of the following reasons: (1) The peak suppression behavior by H is rather different; (2) the entropy obtained from the integration of $C_p/T$ is ~Rln2 corresponding to a doublet ground state of the crystal electric field of a $Ce^{3+}$ ion; and (3) if the heavy electrons give superconductive transitions, the magnitude of the $C_p$ jump should be ~2 J $mol^{-1}$ $K^{-1}$ ($\gamma T_c$); however, the observed jump was ~4 J $mol^{-1}$ $K^{-1}$. This AFM transition has also been pointed out by several groups on the basis of the magnetic susceptibility data on $CeNiBi_2$ [18-20]. A similar small hump near 7 K was shown in the right inset of **Fig. 2(b)**. The evolution of the magnetic peaks for our sample was confirmed below 5 K by powder neutron diffraction [14]. The AFM ordering of the Ce $4f^1$ spin in which the magnetic moment is parallel to the c-axis, was confirmed by the appearance of new magnetic peaks due to the disappearance of n-glide plain.

Mass enhancement presumably results from strong interaction of the Ce 4f electron with the carrier electrons, which may come from Ni 3d electrons mixed with Bi(1) 6p electrons. The peak due to superconductive transition in the specific heat was not observed. The Bi 6p band in the metallic Bi(2) square net contains a positive hole, which may lead to the small blocking effect against the $Ni_xBi(1)$ conducting layer. The mass of carrier in the Bi(2) 6p band is not heavy, that is, the γ value is smaller by an order of magnitude than the observed value of 0.4 J $K^{-2}$ mol-$Ce^{-1}$. The γ value for the Bi square net has not previously been reported. Thus we estimated the γ value from the data on similar compounds. For example, the γ value for the $Ni_{1/3}Bi$ superconductor with $T_c$ = 4 K is reported to be 4.3 mJ $K^{-2}$ $mol^{-1}$, and the specific heat jump, $\Delta C_p$, should be ~$\gamma T_c$, 20 mJ $K^{-1}$ $mol^{-1}$ [21]. The magnitude of this jump relative to the observed AFM peak was too small to be observed. When the light carriers coexisting with the heavy carriers cause superconductivity, the peak originating from the superconductivity is hidden by overlap of the strong peak due to the AFM transition. Therefore, the specific heat data indicated that there are two kinds of carriers with noticeably different effective masses. The heavy carrier is due to magnetic interaction between the conduction electrons and the Ce 4f electron which causes the AFM ordering at ~5 K, and the light carriers cause superconductivity at ~4 K. It is proposed that the origin of the light electrons comes from Bi(2) 6p and superconductivity occurs in the Bi square net, whereas the heavy electrons come from $Ni_xBi(1)$ layers. The $T_c$ (~4 K) for $CeNi_{0.8}Bi_2$ is significantly higher than those of LaNiPO and Ce based mass-enhancement superconductors,[3,22] because the two-dimensional electronic structure weakened by the metallic Bi blocking layer and the mass-enhancement by the $Ce^{3+}$ ion should be disadvantageous for the emergence of superconductivity in the Bi(2) square net.

This work was supported by the Funding Program for World-Leading Innovative R&D on Science and Technology (FIRST), Japan. We thank Drs. S. Shamoto, K. Kodama, and S. Wakimoto (JAER) for powder neutron diffraction measurements.

TABLE I. Lattice constants and unit cell volumes of $RNi_xBi_2$. The lattice constants were determined by LeBail fitting [17].

| R | La | Ce[14] | Nd | Gd[13] | Tb[13] | Dy[13] | Y[13] |
|---|---|---|---|---|---|---|---|
| x | 0.65 | 0.80 | 0.89 | 0.86 | 0.78 | 0.77 | 0.82 |
| a/Å | 4.5599(3) | 4.5439(1) | 4.519(1) | 4.48882(5) | 4.48486(3) | 4.47493(4) | 4.48350(4) |
| c/Å | 9.7544(9) | 9.6414(2) | 9.532(1) | 9.3658(2) | 9.3062(2) | 9.25852(1) | 9.30026(5) |
| Volume/Å$^3$ | 202.75(4) | 199.065(8) | 194.6(2) | 188.715(8) | 187.185(6) | 185.402(6) | 186.952(7) |

TABLE II. Obtained parameters for superconductivity of $RNi_xBi_2$.

| Compound | $T_c$/K | $H_{c1}$/Oe[a] | Volume fraction[a] |
|---|---|---|---|
| $LaNi_{0.65}Bi_2$ | 4.0 | 90 | 0.01 |
| $CeNi_{0.80}Bi_2$ | 4.2 | 65 | 0.96 |
| $NdNi_{0.89}Bi_2$ | 4.1 | 55 | 0.14 |
| $YNi_{0.85}Bi_2$ | 4.1 | 67 | 0.17 |

[a]These were estimated from a M-H curve at 2 K.

**Figures and figure legends**

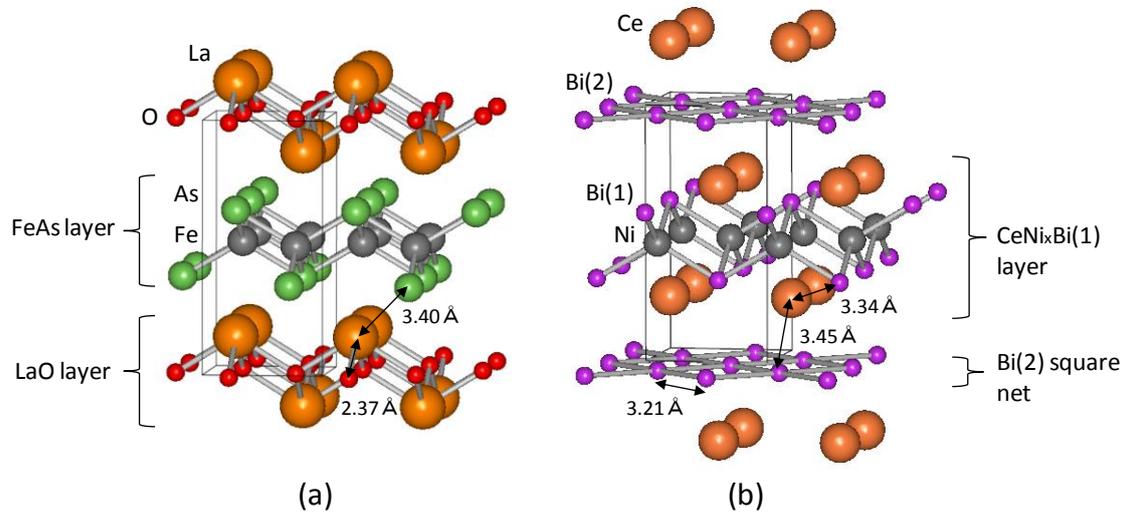

FIG. 1. Crystal structures of (a)LaFeAsO and (b)CeNi$_x$Bi$_2$ belonging to the ZrCuSiAs-type [1, 14]. The valence state of each ion in (b): 3 for Ce, +1 for Ni, −3 for Bi(1) and −1 for Bi(2). Since there is a large valence-difference between Bi(1) and Bi(2), the Ce$^{3+}$ that locates at sites between the two Bi(2)-layers is relaxed to the Bi(1) layers with larger negative charges and the remaining Bi(2) forms a square net.

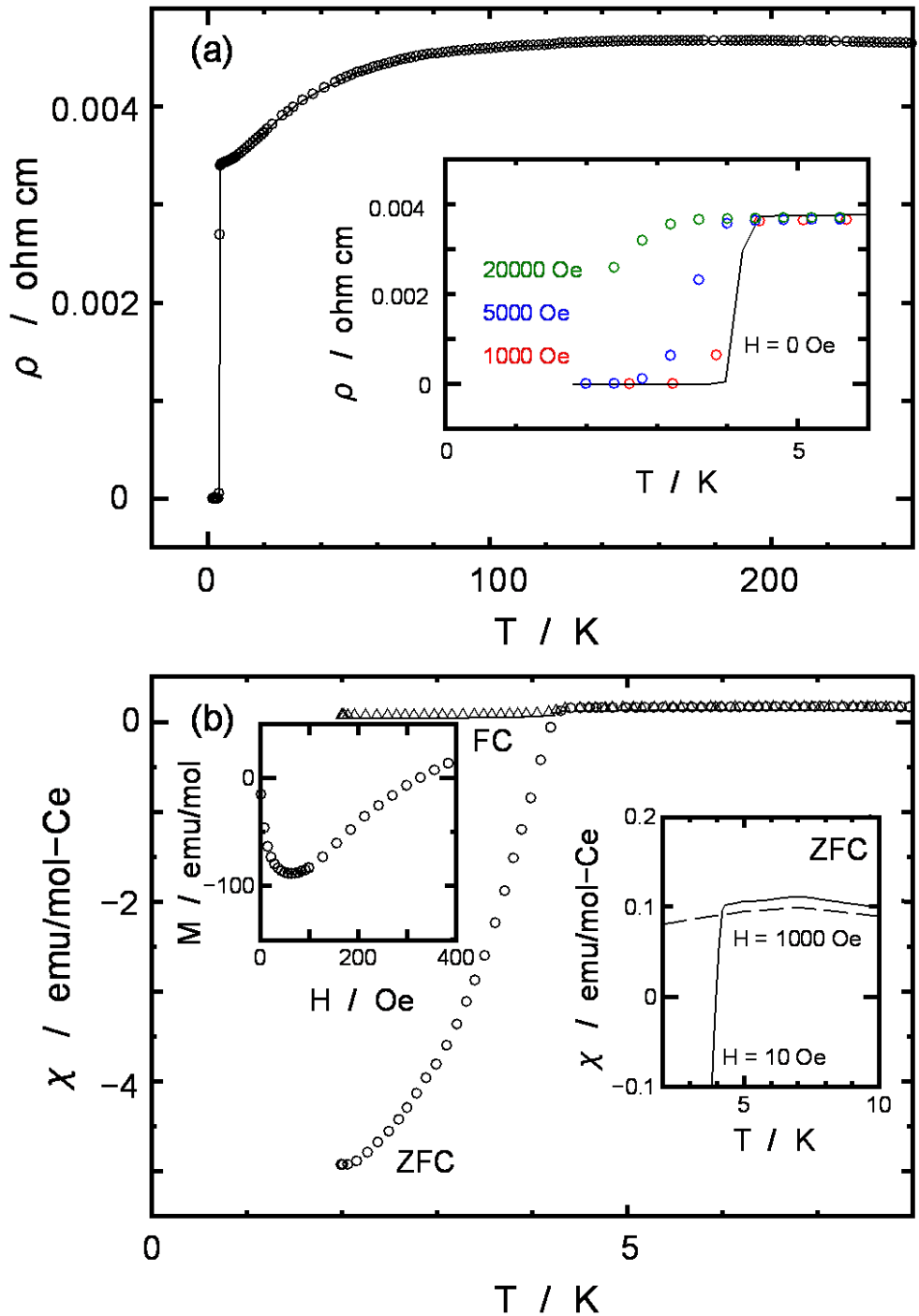

FIG. 2. (a) Temperature dependence of the electrical resistivity (ρ) for CeNi$_{0.8}$Bi$_2$ at 0 Oe. The inset shows the ρ-T curves as a function of the magnetic field. (b) Temperature dependence of the magnetic susceptibility (χ) under conditions of ZFC and FC at 10 Oe. The right inset shows an enlargement around the transition under the condition of ZFC. Two measurement fields, 10 and 1000 Oe were used. The left inset shows the field dependence of the magnetization at 2 K.

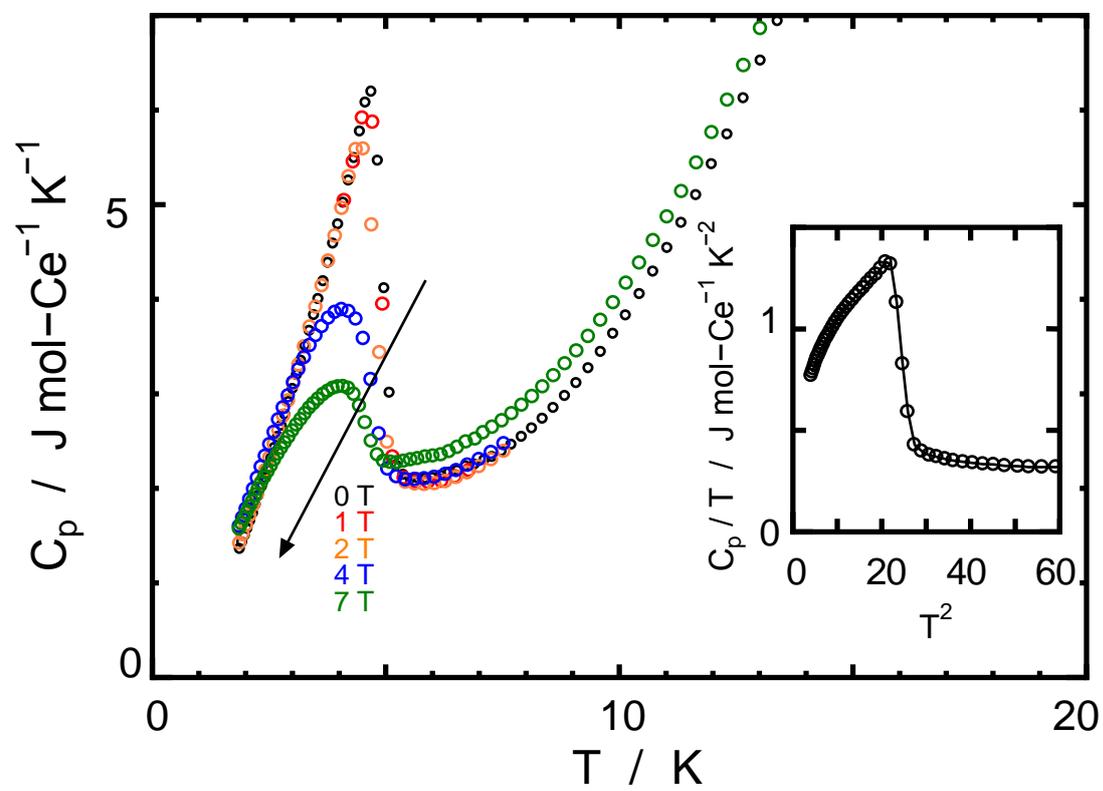

FIG. 3. Magnetic field dependence of the heat capacity of $CeNi_{0.8}Bi_2$ below 15 K. The inset shows the $T^2$-dependence of the heat capacity per unit temperature of $CeNi_{0.8}Bi_2$ at 0 Oe.